# Retinal hazards and glare from space mirrors

Salvador Bará

*Former profesor titular at Universidade de Santiago de Compostela (retired), Santiago de Compostela, 15782 Galicia (Spain, European Union)*

e-mail: salva.bara@usc.gal https://orcid.org/0000-0003-1274-8043

**Abstract:** Projects of orbital mirrors reflecting sunlight towards Earth raise reasonable concerns regarding photochemical and thermal retinal hazards and detrimental effects on visual performance (disability glare). This paper addresses these issues based on some current eye safety and glare standards. Telescopic observations of single mirrors appear to pose a significant risk of photochemical retinal hazard. The risk is higher for vulnerable populations. Unaided eye viewing of a mirror keeps the retinal risks low, both photochemical and thermal; however, disability glare may be unacceptably high, affecting awareness and response in critical visual operations. The simultaneous presence of several mirrors in the visual field of the observer, seen with the unaided eye or through optical instruments, will significantly increase all types of hazards. The kind and severity of their combined effects will depend on the angular distance between them. Additional information is needed to evaluate the effects of the transient flashes of light during mirror slewing operations.

## 1. Introduction

Projects of space mirrors reflecting sunlight towards Earth are not new. From the early works of Hermann Oberth in 1929 to the WWII *Sonnengewehr*, and from NASA studies in the 80s (Canady and Allen 1982) to the partially successful *Znamya* tests in the 90s by the Russian Federal Space Agency, now Roscosmos, they have been a topic deserving intermittent public attention. What once was perceived as a scientific curiosity has become, however, a matter of concern, after the recent authorization granted by the US Federal Communications Commission (FCC) to the firm Reflect Orbital (RO) for deploying a prototype of space mirror in low Earth orbit (LEO), intended to provide sunlight at nighttime with purported military, industrial, energy, agricultural, first response, social, and outdoor lighting applications (FCC 2026).

Although this FCC authorization refers to a single mirror of 18 × 18 m$^2$, the announced final project consists of a macroconstellation of 50 000 space reflectors of 54 × 54 m$^2$, nine times bigger than the first prototype (RO 2026). Understandably, some questions have been raised about the effects of such a project on the quality of the natural night (Zhu and Fu 2026). Sunlight reflected by artificial objects in Earth orbit is just another source of light pollution, a global disruptor of the nighttime environement (Bará and Falchi 2023) with detrimental consequences for science, culture, nature, and, potentially, human health. The sunlight reflected by the cloud of orbital debris already contributes to the increase of the sky brightness baseline (Kocifaj et al 2021; Barentine et al. 2023; Wallner et al. 2026). Large space mirrors are expected to produce significant amounts of skyglow, even at great distances from the illuminated areas, due to the atmospheric scattering of their intense beams of light. Quantitative estimates of this skyglow have been calculated for different scenarios by Hainaut (2026) and Kocifaj et al. (2026).

This paper addresses an additional set of concerns: the consequences of looking at the extremely bright orbital mirrors, either intentionally or by accident, with the unaided eye or with instruments. Our attention will be focused on two main types of hazards: acute retinal damage (mediated by photochemical and thermal interactions) and the impairment of critical vision tasks (disability glare).

The structure of this paper is as follows: in section 2 the main retinal damage mechanisms due to light exposure are briefly recalled. Section 3 discusses the relevance of the angular size of the source for assessing retinal irradiance. Section 4 describes the usual framework for glare evaluation. Section 5 provides basic expressions for calculating the ground radiance and irradiance from orbital mirrors, on which the above effects depend. Quantitative results for retinal hazards and glare with realistic parameters are provided in section 6.

## 2. Thermal and photochemical retinal hazards

Light may damage the retina through photochemical, thermal, and mechanical processes (Glickman 2002). For the radiances and exposure times typical of unaided-eye Sun viewing, thermal and photochemical hazards are dominant. They are quantitatively addressed in the "ICNIRP Guidelines on Limits of Exposure to Incoherent Visible and Infrared Radiation" (ICNIRP 2013), a document that provides the action spectra and the maximum exposure limits for each type of hazard. The reader is referred to it and the references cited threrein for additional details.

Thermal damage of the retina (Vos 1962; White et al. 1971; Schulmeister and Mathieu 2011) is due to local increases of temperature produced by optical radiation in the 380–1400 nm wavelength range. Intraocular media are appreciably transparent in this spectral region, allowing the external radiation to be focused on the retina with little attenuation. Wavelengths outside this range are generally absorbed by tissues of the anterior segment of the eye (cornea, iris, and eye lens).

Photochemical retinal damage known as "Blue-light" or Type II retinopathy (Ham et al. 1976; Ham and Mueller 1989; Lund 2006) is produced by radiation of 380–550 nm for the general adult population, and of 300–550 nm for aphakic people (people whose eye lens was removed without being replaced by an artificial intraocular lens with UV filtering capability) or very young children (because they have extremely transparent eye lenses in the UV-A region). The damage mechanism involves oxidative processes in retinal tissues, especially in the retinal pigment epithelium and the outer segments of photoreceptors (Ham and Mueller 1989; Wu et al. 2006; Roehlecke et al. 2013), which ultimately lead to cell death and loss of visual function. Highly reactive oxidants are produced by the intermediation of photosensitizer molecules which transition to excited states after the absorption of short-wavelenght light (Pattison 2012). Several photosensitizer compounds are naturally produced by the eye itself; other can be present in the retina if administered for photodynamic therapy or other medical treatments (Glickman 2002; Godley et al. 2005; Arnault et al. 2013).

During direct solar observations both damage mechanisms are at play, and may act synergistically. However, contrary to an extended popular belief, the main cause of solar retinopathy when looking at the Sun with the unaided eye and normally constricted pupils is not thermal, but photochemical (van Norren and Vos 2016). In Ham's words, "(...) solar retinitis and eclipse blindness are photochemical phenomena and near infra-red solar radiation makes only a negligible contribution to retinal damage" (Ham et al. 1980). Keep in mind, however, that this does not automatically apply to observations made with optical instruments (Section 3.3).

The maximum exposure limits set forth in ICNIRP (2013) are expressed, depending on the angular size of the source, in terms of different radiometric quantities. For sources of large angular extent, effective radiances L (W m$^{-2}$ sr$^{-1}$) and radiance doses D (J m$^{-2}$ sr$^{-1}$) are used by default. For small sources, the limits can be equivalently expressed in terms of irradiances E (W m$^{-2}$) and radiant exposures H (J m$^{-2}$), which in practice are easier to measure than radiances. All these quantities are evaluated on a plane tangent to the eye cornea and perpendicular to the line of sight, by spectrally weighting the incident radiation with the action spectrum corresponding to each hazard mechanism (Fig. 1).

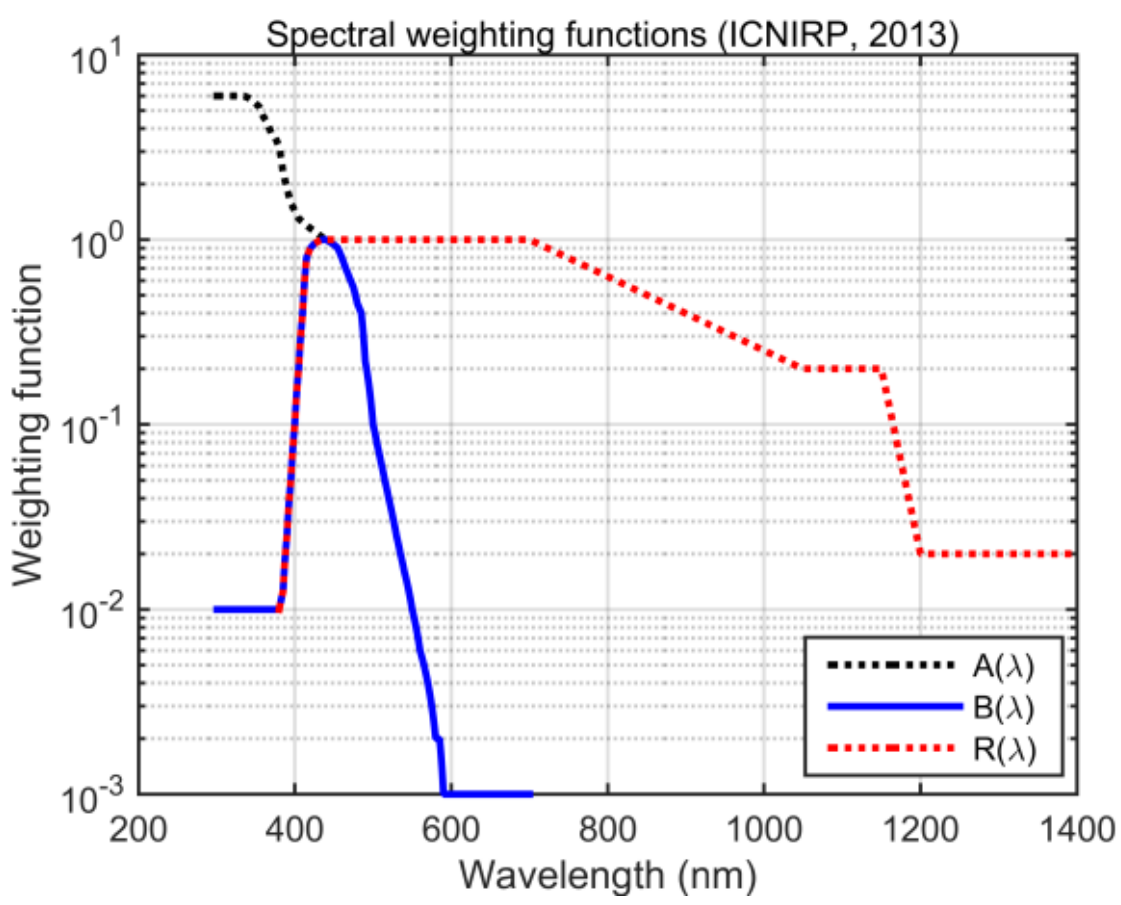

**Fig 1.** Weighting functions for the evaluation of retinal hazards (ICNIRP 2013). A($\lambda$) Blue light retinopathy aphakic population; B($\lambda$) Blue light retinopathy, general population; R($\lambda$) Retinal thermal hazard.

A relevant output from retinal safety assessments is the amount of time required to reach the maximum exposure limit. If this time is too short, natural defense mechanisms like the aversion reflex (blinking, changing the line of sight,...) may not be sufficient to prevent harm. Detailed measurement methods for evaluating the exposure to light from indoor and outdoor lamps are available in IEC (2013), which also provides a lamp classification scheme in four classes of risk, assuming a well-functioning aversion response. With suitable adaptations, a similar type of classification could be adopted for orbital mirrors. Note however that ocular damage can be produced by a defective or intentionally overridden aversion reflex, as it has been reported in solar eclipse viewing without sufficient protection (cfr., inter alia, Dhir et al. 1981; Michaelides et al. 2001; Wong et al. 2001; Moran and O'Donoghue 2013; Gregory-Roberts et al. 2015) and inadequate use of laser pointers (Luttrull and Hallisey 1999; Sell and Bryan 1999; Zamir 1999; McGhee et al. 2000; Farajpour et al. 2015; van Norren and Vos 2016). In the present study we will use as relevant output the time to reach the maximum exposure limits, leaving for future works the development of a specific mirror risk classification table.

## 3. The relevance of the angular size of the source for assessing retinal exposure

The intensity of the retinal effects of optical radiation depends, among other factors, on the angular extent of the light source, viewed from the observer location. The standard corneal exposure limits (ICNIRP 2013) contemplate explicitly this parameter.

To better understand the role of the angular size of the source, it is useful to see how does it appear in the expressions for the retinal spectral irradiance, the radiometric quantity that determines the amount of energy reaching the retinal cells per unit area, unit time, and unit wavelength interval. The total retinal irradiance is obtained by integrating the spectral retinal irradiance across the range of wavelengths (λ) of the illuminating light. The retinal radiant exposure is finally obtained by adding up the retinal irradiance during the exposure time.

### *3.1. Geometrical factors*

To formalize it, let us consider a spatially homogeneous source of spectral radiance $L_{s,\lambda}$ (W m$^{-2}$ sr$^{-1}$ nm$^{-1}$), evaluated at the the eye cornea, and subtending a solid angle $\Omega_s$ (sr) seen from the observer[1]. The spectral irradiance $E_{s,\lambda}$ on a plane tangent to the cornea and perpendicular to the line of sight is $E_{s,\lambda} = L_{s,\lambda}\,\Omega_s \cos\theta$ (W m$^{-2}$ nm$^{-1}$), where $\theta$ is the angle between the source direction and the line of sight. Henceforth we will assume that the observer is looking directly at the source, so that $\cos\theta = 1$. Otherwise, the factor $\cos\theta$ shall be included in the final results.

If the diameter of the eye pupil is $d_p$ (m), the spectral radiant flux entering the eye is $\Phi_{s,\lambda} = E_{s,\lambda}\,\pi\,(d_p/2)^2 = \pi\,(d_p/2)^2 L_{s,\lambda}\,\Omega_s$ (W nm$^{-1}$). The spectral radiant flux arriving to the retinal layers is $\Phi_{r,\lambda} = \xi_\lambda\,\Phi_{s,\lambda}$ where $\xi_\lambda$ is the spectral transmittance of the ocular media (we use here the Greek letter $\xi$ for intraocular transmittance instead of the $\tau$ used in ICNIRP 2013, to avoid confusion with the same symbol used for atmospheric molecular and aerosol optical depths in

[1] If the source has circular shape, with diameter $\alpha_s$ (in radian), then $\Omega_s = 2\pi[1-\cos(\alpha_s/2)]$ sr, which for $\alpha_s \ll 1$ rad reduces to $\Omega_s \approx \pi\,(\alpha_s/2)^2$ sr. Under the same conditions, a square source of side $\alpha_s$ woud subtend a solid angle $\Omega_s \approx {\alpha_s}^2$ sr.

Section 6). The retinal irradiance $E_{\mathrm{r},\lambda}$ is the ratio of the retinal radiant flux $\Phi_{\mathrm{r},\lambda}$ to the area of the retinal image of the source, $A_{\mathrm{r}}$:

$$E_{\mathrm{r},\lambda} = \frac{\Phi_{\mathrm{r},\lambda}}{A_{\mathrm{r}}} = \frac{\pi\, \xi_\lambda\, {d_{\mathrm{p}}}^2}{4\, A_{\mathrm{r}}} L_{\mathrm{s},\lambda}\, \Omega_{\mathrm{s}} \tag{1}$$

And in evaluating $A_{\mathrm{r}}$ is where the angular size of the source relative to the size of the point spread function of the eye (PSF, the retinal image of a point source, Goodman JW 1996) becomes a relevant factor in real eyes. This relevance is further reinforced by the continuous and involuntary eye movements, and by the heat transport properties of the retina, as described below (Sect. 3.2).

If the eye optics were ideal, in the geometrical optics sense, the PSF of the eye would be a Dirac-delta distribution (i.e. the image of any mathematical point would be another mathematical point) and the lateral magnification of the eye would be constant, so that, excepting for an overall magnification factor, the retinal image would be an exact replica of the source. The area of the retinal image would be $A_{\mathrm{r}} = A_{\mathrm{i}} = f^2\, \Omega_{\mathrm{s}}$ ($\mathrm{m}^2$) where $A_{\mathrm{i}}$ is the area of the ideal image of the source, and $f$ is the equivalent focal length of the eye in air (conventionally taken as 16.67 mm). Under these conditions Eq. (1) becomes:

$$E_{\mathrm{r},\lambda}(\text{ideal eye}) = \frac{\pi\, \xi_\lambda\, {d_{\mathrm{p}}}^2}{4\, f^2} L_{\mathrm{s},\lambda} \tag{2}$$

(see also Eq. 2 in ICNIRP, 2013). Equation (2) shows that in an ideal eye the retinal irradiance depends on the radiance at the cornea, but not on the corneal irradiance or the angular size of the source.

However, actual eyes are far from ideal in the geometrical optics sense. The PSF of a physiologically healthy human eye is not a mathematical point but a blurred spot of light of finite size (Navarro et al. 1998). This is due to the combined action of difraction from the finite pupil aperture (Born and Wolf 1999), uncompensated ametropies (myopia, hyperopia, astigmatism), high-order eye aberrations (Thibos et al. 2002), and intraocular scattering (van den Berg 2010, 2013). As a result, real retinal images are blurred versions of the original objects. In mathematical terms, the actual image is given by the two-dimensional convolution of the ideal geometric image with the PSF of the eye (Goodman JW 1996). Since the linear dimension of a convolution is the sum in quadrature of the linear dimensions of the intervening functions (this is strictly valid only for Gaussian distributions, but approximately valid for many real image profiles), the area of the actual image ($A_{\mathrm{r}}$) is the sum of the areas of the ideal image ($A_{\mathrm{i}}$) and the PSF ($A_{\mathrm{PSF}}$). For actual eyes, then, Eq. (1) has the general form:

$$E_{\mathrm{r},\lambda}(\text{actual eye}) = \frac{\pi\, \xi_\lambda\, {d_{\mathrm{p}}}^2}{4\,(A_{\mathrm{i}} + A_{\mathrm{PSF}})} L_{\mathrm{s},\lambda}\, \Omega_{\mathrm{s}} = \frac{\pi\, \xi_\lambda\, {d_{\mathrm{p}}}^2}{4\, f^2 \left(1 + \frac{\Omega_{\mathrm{PSF}}}{\Omega_{\mathrm{s}}}\right)} L_{\mathrm{s},\lambda} \tag{3}$$

where $A_{\mathrm{PSF}} = f^2\, \Omega_{\mathrm{PSF}}$, being $\Omega_{\mathrm{PSF}}$ the solid-angle subtended by the PSF back-projected into the object space. Note that when the ideal image of the source is much larger than the PSF ($\Omega_{\mathrm{PSF}}/\Omega_{\mathrm{s}} \ll 1$) Eq.(3) converges to Eq. (2) and the actual eye behaves in an approximately ideal way. However, when the source is smaller than the PSF ($\Omega_{\mathrm{PSF}}/\Omega_{\mathrm{s}} \gg 1$), Eq.(3) becomes:

$$E_{\mathrm{r},\lambda}(\text{actual eye, small source}) = \frac{\pi\, \xi_\lambda\, {d_\mathrm{p}}^2}{4\, f^2\, \Omega_\mathrm{PSF}} E_{\mathrm{s},\lambda} \tag{4}$$

which allows to express the exposure limits for small sources in terms of the incident corneal irradiance, $E_{\mathrm{s},\lambda}$. Equation (4) basically accounts for the fact that the radiant flux entering the eye is spread over the PSF area, which is in this case much larger than the geometrical image of the source.

It is worth noting that if Eq. (2) were used instead of the correct Eq. (3) to calculate the retinal irradiance produced by a small source on a static real eye ($\Omega_\mathrm{PSF}/\Omega_\mathrm{s} \gg 1$), the result would be grossly overestimated by a factor of order $\Omega_\mathrm{PSF}/\Omega_\mathrm{s}$. This follows from substituting the definition $E_{\mathrm{s},\lambda} = L_{\mathrm{s},\lambda}\, \Omega_\mathrm{s}$ in the correct expression for this condition, Eq. (4). To get some insight about the magnitude of this overestimation, compare the linear angular extent of an orbital mirror, ~0.3 arcmin (for a 54 m wide mirror at 625 km altitude), with the angular diameter of the Sun (~30 arcmin), and a typical human PSF (a few arcmin). The corresponding solid angles are proportional to the squares of these quantities. Whereas $\Omega_\mathrm{PSF} \ll \Omega_\mathrm{SUN}$ ($\Omega_\mathrm{PSF}/\Omega_\mathrm{SUN} \approx 0.01$), and hence Eq. (2) applies, we have $\Omega_\mathrm{PSF} \gg \Omega_\mathrm{s}$ ($\Omega_\mathrm{PSF}/\Omega_\mathrm{s} \approx 100$), for which Eq.(4) should be used instead. That's why back-of-the-envelope estimations of the retinal exposure from orbital mirrors, by comparing them with the Sun exposure evaluated via Eq.(2), and based on the similarity of their exceedingly high luminances, $L_\mathrm{s}{\sim}10^9$ cd/m$^2$ (sect. 5), may easily lead to erroneous conclusions.

### *3.2. Time matters: eye movements and thermal transport*

Time is an essential parameter when it comes to evaluating the received dose. The angular size of the source also plays a relevant role here.

The eye, even when intentionally fixating on a static target, is in continuous movement. The instantaneous line of sight fluctuates randomly around its average direction, with three main components: *drifts*, which are slow movements with typical amplitudes of 1–9 arcmin, *microsaccades* occurring 2–10 times per second, with ~25 millisecond duration and typical amplitudes in the range 13–67 arcmin, and *tremors*, movements of very low amplitude (0.06–0.48 arcmin) but with frequencies in the range 50–100 Hz (Abadi and Gowen 2004; Møller et al. 2006; Arines et al. 2009). These involuntary movements are instrumental for vision, since they contribute to avoid a steady, constant signal on the individual retinal photoreceptors, whose perception would be quicky suppressed by the visual system.

The image of a light source moves on the retina according to these components. If the angular extent of the source is small, this leads to a smaller time-averaged irradiance in the exposed area than that deduced from Eq. (1). The effect would be equivalent to increasing the size of $A_\mathrm{r}$, accounting for the area effectively irradiated during the considered period of time due to the eye movements. The exposure limits in ICNIRP (2013) and derived works already include the correction factors to take into account this effect.

Thermal damage is also dependent on the angular size of the source. Part of the heat produced by absorption of light at any point of the retina is carried away by thermal diffusion in the retinal tissue and by blood transport along the retinal capillary network. This effect is particularly noticeable for small-sized retinal images and exposure times allowing thermal transport processes. Heat drain makes the retina more resilient to thermal exposure, everything else being equal. As in the case of involuntary eye movements, this effect is explictly taken into account in the ICNIRP (2013) exposure limits.

### *3.3. Observing with telescopes*

When the source is observed through a telescope, the radiometric quantities at the entrance of the eye, the effective pupil size, and the size of the retinal image are modified, with important consequences for retinal hazard assessment. To formalize it, let us consider a telescope of diameter $D_{\mathrm{T}}$, working with angular magnification $\beta$, and internal spectral transmittance $\xi_{T,\lambda}$. In the usual observing configuration, its exit pupil, of diameter $d_{\mathrm{T}} = D_{\mathrm{T}}/\beta$ , lies in the same plane as the eye input pupil, of diamenter $d_{\mathrm{p}}$, and it is concentric with it. The pupil sizes may often be unmatched, $d_{\mathrm{T}} \neq d_{\mathrm{p}}$, so that the effective pupil of the eye is the minimum of $(d_{\mathrm{p}}, d_{\mathrm{T}})$.

Due to the internal transmittance of the optics, the spectral radiance at the exit pupil of the telescope becomes $L_{\mathrm{s},\lambda}{}' = \xi_{T,\lambda}\, L_{\mathrm{s},\lambda}$, a slightly attenuated version of the $L_{\mathrm{s},\lambda}$ incident on its input pupil. The linear angular size of the source, viewed through the instrument, becomes $\alpha_{\mathrm{s}}{}' = \beta\ \alpha_{\mathrm{s}}$ rad, subtending a correspondingly larger solid angle $\Omega_{\mathrm{s}}' = \beta^2\ \Omega_{\mathrm{s}}$ sr. The spectral irradiance at the exit pupil is then $E_{\mathrm{s},\lambda}' = L_{\mathrm{s},\lambda}'\Omega_{\mathrm{s}}' = \beta^2\ \xi_{T,\lambda}\ L_{\mathrm{s},\lambda}\ \Omega_{\mathrm{s}} = \beta^2\ \xi_{T,\lambda}\ E_{\mathrm{s},\lambda}$. The diameter of the effective pupil at the entrance of the eye will be $d_{\mathrm{p}}' = \min(d_{\mathrm{p}}, d_{\mathrm{T}})$ .

The spectral retinal irradiance for telescopic viewing is then given by Eqs. (1)–(4) with the new values at the entrance of the eye, that is, using $L_{\mathrm{s},\lambda}{}'$, $\Omega_{\mathrm{s}}{}'$, $E_{\mathrm{s},\lambda}{}'$, and $d_{\mathrm{p}}{}'$, instead of $L_{\mathrm{s},\lambda}$, $\Omega_{\mathrm{s}}$, $E_{\mathrm{s},\lambda}$, and $d_{\mathrm{p}}$, respectively. Equation (3), e.g., becomes:

$$E_{\mathrm{r},\lambda}(\text{actual eye, with telescope}) = \frac{\pi\ \xi_{\lambda}\xi_{T,\lambda}\left[\min\left(d_{\mathrm{p}}, D_{\mathrm{T}}/\beta\right)\right]^2}{4\,f^2\left(1 + \frac{\Omega_{\mathrm{PSF}}}{\beta^2\Omega_{\mathrm{s}}}\right)} L_{\mathrm{s},\lambda} \tag{5}$$

Note that when observing with a magnifying telescope ($\beta \geq 1$) with matched pupils ($d_{\mathrm{p}} = d_{\mathrm{T}} = D_{\mathrm{T}}/\beta$) an extended source, i.e. a source which is already well resolved with the unaided eye ($\Omega_{\mathrm{PSF}}/\Omega_{\mathrm{s}} \ll 1$), the retinal irradiance does not increase (in fact it becomes slightly smaller, due to the transmittance $\xi_{T,\lambda} \leq 1$). However, the retinal risk can be significantly larger in case of telescopic observations, due the substantial increase in the exposed retinal area (Schulmeister 2013), which is $\beta^2$ times larger.

The increase of risk is more noticeable in case of small, unresolved sources ($\Omega_{\mathrm{PSF}}/\Omega_{\mathrm{s}} \gg 1$). If they continue to be small in telescopic viewing, i.e., if $\Omega_{\mathrm{PSF}}/(\beta^2\Omega_{\mathrm{s}}) \gg 1$, the retinal irradiance increases by a factor $\beta^2$, because the large radiant flux captured by the telescope aperture, which is $\beta^2$ times larger compared with the unaided eye and is fed entirely into it under the matched pupil sizes condition, is distributed over an essentially constant retinal area ($\approx A_{\mathrm{PSF}}$). Alternatively, for sufficiently large solid angle magnifications $\beta^2$, small unaided eye sources ($\Omega_{\mathrm{PSF}}/\Omega_{\mathrm{s}} \gg 1$) may become large telescopic ones ($\Omega_{\mathrm{PSF}}/(\beta^2\Omega_{\mathrm{s}}) \ll 1$), losing the resilience provided by PSF blur, eye movements and thermal drainage (section 3.2.).

Observing bright sources (e.g. the Sun) through an unfiltered telescope poses a high risk of another type of hazard, not included in the above analyses: immediate thermal burns of the cornea. Recall that the irradiance at the exit pupil of the telescope is $\beta^2\, \xi_{T,\lambda}$ times larger than that of the natural Sun. Observing the Sun with an unfiltered telescope ($\xi_{T,\lambda} \approx 1$) at usual magnifications of order $\beta{\sim}100$ would amount to increasing by a factor 10000 the Sun irradiance on the cornea, with dramatic consequences for this and other tissues of the anterior segment of the eye (iris, eye lens).

As a final remark, unaided eye observations can be formally considered a particular case of Eq.(5), in which the telescope works at magnification $1\,\times$, has a diameter $D_{\mathrm{T}}$ equal to or larger than the diameter of the eye pupil, and is perfectly non-attenuating, $\xi_{T,\lambda} = 1$, leading to Eq.(3).

## 4. Glare

Light scattered within the eye impairs vision performance by reducing the contrast between the objects and their surrounding field. From a perceptual viewpoint this effect is known as disability glare. All human eyes are affected by it, to a greater or lesser extent, due to the unavoidable inhomogeneities of ocular tissues at different size scales. Other factors being equal, intraocular scattering tends to increase with age and is more severe in weakly pigmented eyes (van den Berg 2010, 2013).

Since the visual detection of objects largely depends on them having enough luminance contrast against the surrounding field (Blackwell 1946), an increase of glare, especially if it is sudden and unexpected, poses a relevant risk for critical vision tasks, particularly for drivers of vehicles, including ships and airplanes. Glare also makes it more difficult to see celestial objects with the unaided eye, because it adds to artificial skyglow in the retina (Bará and Bao-Varela 2023).

The photometric magnitudes used to quantify glare are the luminance (cd m$^{-2}$) and the illuminance (lx). Recall that the luminance, $L_{\mathrm{v}}$, is just the integral over wavelengths of the spectral radiance $L_{\mathrm{s},\lambda}$, weighted by the spectral luminous efficiency for photopic vision, $V_\lambda$ (CIE 2019), with a global scale factor $K = 683$ lm/W. The integral is extended to the visible spectrum, and can be approximated by a finite sum over wavelength intervals, $\Delta\lambda_i$, as shown in Eq. (6):

$$L_{\mathrm{v}} = K \int_{\lambda=380\ \mathrm{nm}}^{\lambda=740\ \mathrm{nm}} V_\lambda\, L_{\mathrm{s},\lambda}\, \mathrm{d}\lambda \;\cong\; K \sum_{i=1}^{I} V_{\lambda_i}\, L_{\mathrm{s},\lambda_i}\, \Delta\lambda_i \tag{6}$$

A similar equation holds for the illuminance $E_{\mathrm{v}}$, using the spectral radiance $E_{\mathrm{s},\lambda}$ instead of $L_{\mathrm{s},\lambda}$.

A specific glare descriptor is the equivalent veiling luminance $L_{\text{veil}}$ (cd m$^{-2}$), that is, the external luminance that would produce in a perfect, non-scattering eye the same visual effect as the considered level of glare in the real eye. The foveal veiling luminance for a point source located at an angle $\theta$ from the line of sight is modeled as

$$L_{\text{veil}}(\theta) = \Psi(\theta)\, E_{\text{glare}} \qquad (7)$$

where $E_{\text{glare}}$ is the illuminance (lx) produced by that source on the eye pupil and $\Psi(\theta)$ is a PSF-like function with dimensions sr$^{-1}$. Different versions of $\Psi(\theta)$ are available in the literature, depending on the range of $\theta$ and on additional parameters like the age and the eye pigmentation of the observer. The *general glare equation* (van den Berg 2010), valid for angles $\theta$ from 0.1° to 100°, is:

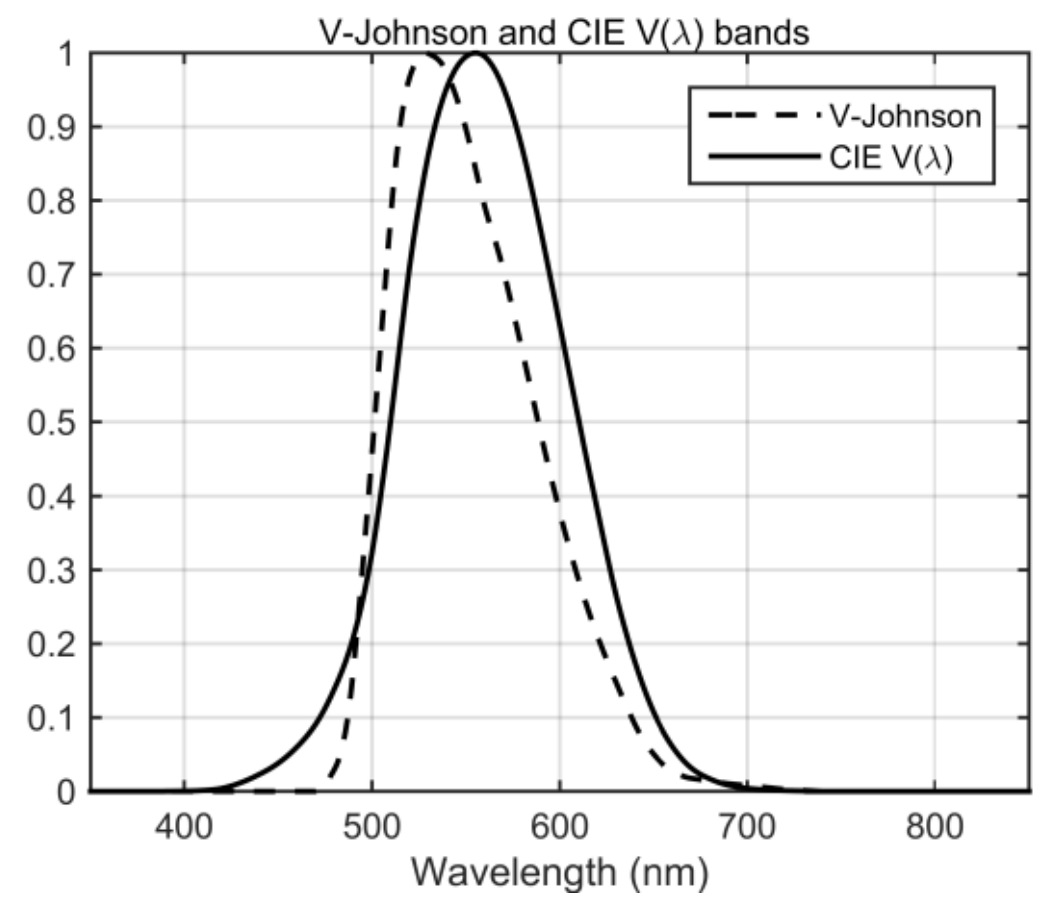

**Fig. 2.** Spectral luminous efficiency for photopic vision, $V_{\lambda}$ (CIE 2019) and Johnson V band.

$$\Psi_{\text{G}}(\theta) = \frac{10}{\theta^3} + \left(\frac{5}{\theta^2} + \frac{0.1 \cdot p}{\theta}\right)\left[1 + \left(\frac{A_y}{62.5}\right)^4\right] + 2.5 \cdot 10^{-3} \cdot p \quad (\text{sr}^{-1}) \qquad (8)$$

where $\theta$ is expressed in degrees (°), $A_y$ is the observer's age (in years) and $p$ is an iris pigmentation factor which takes values from 0 (strongly pigmented eyes) to 1 (weakly pigmented ones).

The EN 13201-3:2015 norm on road lighting (CEN 2015) evaluates glare assuming that all intervening sources are point-like, using Eq. (8) with $p = 0$ for angles $0.1° < \theta \leq 1.5°$, and Eq. (9) for angles $1.5° < \theta \leq 60°$ (light sources located at angles $\theta$ larger than 60° are not taken into account in this norm):

$$\Psi_{\text{S}}(\theta) = \frac{9.86}{\theta^2}\left[1 + \left(\frac{A_y}{66.4}\right)^4\right] \qquad (\text{sr}^{-1}) \qquad (9)$$

Eq. (9) is a modified version of the classical age-adapted Stilles-Holladay equation (van den Berg 2010), with slightly different parameters to extend its angular range of validity. The age of the observer suggested in the norm for calculations is 23 yr (the median age in the world being 31.1 yr). It is worth noting that (i) the use of two different expressions in EN 13201-3:2015 arises from an update of the previous version of the norm, in order to encompass angles smaler than 1.5°, (ii) the resulting, combined PSF is not continuous at the contact point $\theta = 1.5°$, and (iii) the parameters age and pigmentation suggested or chosen in the norm underestimate the amount of glare in the median population.

The severity of the disability glare is evaluated in EN 13201-3:2015 by means of the 'threshold increment', $f_{\text{TI}}$, expressed in %. This metric is defined as "the percentage increase in luminance contrast threshold required between an object and its background for it to be seen equally well with a source of glare present" (CIE 2020). Recall that an object of luminance $L_{\mathrm{O}}$ seen against a background of luminance $L_{\mathrm{B}}$ has a Weber constrast $\gamma_{\mathrm{W}} = (L_{\mathrm{O}} - L_{\mathrm{B}})/L_{\mathrm{B}}$. For the object to be just detectable, $\gamma_{\mathrm{W}}$ has to be equal to the corresponding luminance contrast threshold, $\gamma_{\mathrm{W,0}}$. This threshold depends on the luminance the observer is adapted to, and on the angular size of the object (Blackwell 1946). Intraocular scattering adds a veiling luminance $L_{\mathrm{veil}}$ to both, object and background, such that their new contrast $\gamma_{\mathrm{W}}{}' = (L_{\mathrm{O}} - L_{\mathrm{B}})/(L_{\mathrm{B}} + L_{\mathrm{veil}})$ falls below threshold, $\gamma_{\mathrm{W}}{}' < \gamma_{\mathrm{W,0}}$, and the object can no longer be detected. The object should then have an initial contrast $\gamma_{\mathrm{W}}$ larger than $\gamma_{\mathrm{W,0}}$ (in absence of glare) to be just detectable under this amount of glare.

The threshold increment due to a set of $K$ light sources in the visual field of the observer is calculated as:

$$f_{\mathrm{TI}} = \frac{65}{(L_{\mathrm{av}})^{0.8}} \sum_{k=1}^{K} \Psi(\theta_k)\, E_{\mathrm{v}k} \quad (\%) \tag{10}$$

where $L_{\mathrm{av}}$ is the average background luminance (cd m$^{-2}$), $\theta_k$ (°) is the angle between the direction of the $k$-th source and the line of sight, and the product $\Psi(\theta_k)\, E_{\mathrm{v}k}$ is $L_{\mathrm{veil}}(\theta_k)$, Eq. (7), the veiling luminance produced by the $k$-th source. $E_{\mathrm{v}k}$ is the illuminance (lx) produced by that source on a plane tangent to the eye cornea and perpendicular to the line of sight (recall that $E_{\mathrm{v}k} = E_{\mathrm{v}k}^{\perp} \cos\theta_k$, where $E_{\mathrm{v}k}^{\perp}$ is the illuminance of the source on a plane tangent to the cornea and perpendicular to the source direction $\theta_k$), and $\Psi(\theta_k)$ (sr$^{-1}$) is given by Eq. (8) or Eq. (9), depending on the value of $\theta_k$. Current regulations for road lighting tend to reject installations whose $f_{\mathrm{TI}}$ is larger than 10%-15%.

## 5. Ground radiance and irradiance from orbital mirrors

As follows from the equations in Sections 3 and 4, the spectral radiance $L_{\mathrm{s},\lambda}$ or the irradiance $E_{\mathrm{s},\lambda}$ at the eye's pupil when the observer is fixating on the mirror ($\theta = 0°$) are the key radiometric quantities to evaluate retinal photochemical and thermal hazards. Glare is evaluated in terms of the $f_{\mathrm{TI}}$, which depends on the illuminance on the pupil, $E_{\mathrm{v}}$, calculated in turn from the spectral irradiance $E_{\mathrm{s},\lambda} \cos\theta$ produced by an orbital mirror located at an angle $\theta$ with respect to the line of sight. The problem reduces then to the determination of $L_{\mathrm{s},\lambda}$ and $E_{\mathrm{s},\lambda}$.

At the most basic description level, an orbital mirror works as a pinhole camera projecting on Earth an image of the Sun disc. Pinhole cameras working by reflection (i.e. with a small mirror instead of a pinhole) are traditional devices for solar imaging during eclipses and planetary transits (Wood 1934; Cumming et al. 2024; Rapson et al. 2025). The projected image is an attenuated replica of the Sun luminance distribution, including the edge darkening of the Sun disc. The image is circular in the transversal plane perpendicular to the line joining the pupil of the observer and the satellite, and it becomes elliptical when projected on ground.

Using this basic model, in the geometrical optics approximation, the evaluation of $L_{\mathrm{s},\lambda}$ and $E_{\mathrm{s},\lambda}$ is immediate. Let us consider a space mirror of linear dimension $l_{\mathrm{m}}$ and area $A_{\mathrm{m}}$ orbiting the Earth at an altitude $h_{\mathrm{m}}$ and at a distance $d_{\mathrm{m}}$ to the observer who receives its light. The mirror is oriented such that the angle between its normal and the Sun direction is $\varphi$. The satellite is located at a zenith angle $z$ in the observer's reference frame. Denoting by $E_{\odot,\lambda}$ the spectral irradiance of the Sun at 1 astronomical unit of distance (1 au = 149 597 870 700 m) and by $d_{\odot}$ the distace Sun-satellite, the spectral radiant flux captured by the mirror $\Phi^{\mathrm{in}}_{\mathrm{m},\lambda}$ (W nm$^{-1}$) is:

$$\Phi^{\mathrm{in}}_{\mathrm{m},\lambda} = E_{\odot,\lambda}\left(\frac{1\ \mathrm{au}}{d_{\odot}}\right)^2 A_{\mathrm{m}}\cos\varphi \tag{11}$$

The reflected radiant flux is $\Phi^{\mathrm{out}}_{\mathrm{m},\lambda} = \rho_\lambda\,\Phi^{\mathrm{in}}_{\mathrm{m},\lambda}$, where $\rho_\lambda$ is the spectral reflectance of the mirror. The fraction of this flux reaching the Earth surface is $\Phi^{\mathrm{g}}_{\mathrm{m},\lambda} = \xi_{\mathrm{ATM},\lambda}\,\Phi^{\mathrm{out}}_{\mathrm{m},\lambda}$, where $\xi_{\mathrm{ATM},\lambda}$ is the spectral atmospheric transmittance given by:

$$\xi_{\mathrm{ATM},\lambda} = \exp[-\tau_\lambda\, M(z)] \tag{12}$$

where $\tau_\lambda$ is the atmospheric spectral optical depth, sum of the Rayleigh $\tau_{\mathrm{R},\lambda}$ and aerosol $\tau_{\mathrm{A},\lambda}$ contributions, and $M(z)$ is the airmass number. For satellites located not too close to the horizon (zenith angles $z$ not too close to 90°) the approximation $M(z) \approx 1/\cos z$ provides adequate results. A general expression valid for zenith angles up to 90° is given by Kasten and Young (1989):

$$M(z) = 1/[\cos z + 0.50572 \times (96.07995° - z°)^{-1.6364}] \tag{13}$$

The spectral radiant flux arriving to the observer, $\Phi^{\mathrm{g}}_{\mathrm{m},\lambda}$, is spread over the area $A_{\odot,\mathrm{g}}$ of the projected Sun image. In the plane tangent to the observer's cornea and perpendicular to the line joining the observer pupil and the satellite this area is circular, of diameter $(1\ \mathrm{au}/d_{\odot})\ \alpha_{\odot}\ d_{\mathrm{m}}$, where $\alpha_{\odot}$ (rad) is the angular diameter of the Sun at 1 au. The area of this image is[2]:

$$A_{\odot,\mathrm{g}} = \Omega_{\odot}\left(\frac{1\ \mathrm{au}}{d_{\odot}}\right)^2 {d_{\mathrm{m}}}^2 \tag{14}$$

where $\Omega_{\odot} = 2\pi[1-\cos(\alpha_{\odot}/2)] \cong \pi(\alpha_{\odot}/2)^2$ (sr) is the solid angle subtended by the Sun at 1 au.

Assuming for simplicity that the radiant flux is evenly distributed in the projected image, i.e., neglecting the darkening of the Sun disk towards the edges, the spectral irradiance at the observer's pupil is, then:

$$E_{\mathrm{s},\lambda} = \frac{\Phi^{\mathrm{g}}_{\mathrm{m},\lambda}}{A_{\odot,\mathrm{g}}} = \left(\frac{E_{\odot,\lambda}}{\Omega_{\odot}}\right)\left(\frac{A_{\mathrm{m}}\cos\varphi}{{d_{\mathrm{m}}}^2}\right)\rho_\lambda\exp[-\tau_\lambda\, M(z)] \tag{15}$$

which is explicitly written as the product of three terms: the extraatmospheric spectral radiance of the Sun ($L_{\odot,\lambda} = E_{\odot,\lambda}/\Omega_{\odot}$), the solid angle subtended by the space mirror seen from the observer location ($\Omega_{\mathrm{m}} = A_{\mathrm{m}} \cos\varphi / \, {d_{\mathrm{m}}}^2$), and the spectral transmittance of the overall light path ($\rho_\lambda \exp[-\tau_\lambda \, M(z)]$). The spectral radiance at the eye pupil can be calculated as:

$$L_{\mathrm{s},\lambda} = \frac{E_{\mathrm{s},\lambda}}{\Omega_{\mathrm{m}}} = L_{\odot,\lambda} \, \rho_\lambda \exp[-\tau_\lambda \, M(z)] \qquad (16)$$

a result that could be directly deduced from the theorem of invariance of radiance, whereby the radiance is conserved along geometrical rays excepting for intrinsic attenuation in material media. The radiance at the observer, $L_{\mathrm{s},\lambda}$, is then the extraatmospheric radiance of the Sun, $L_{\odot,\lambda}$, slightly reduced by the mirror reflectance and the attenuation along the atmospheric path.

[2] Strictly speaking, the projected image would be the convolution of the ideal projected image of the Sun (of area $A_{\odot,\mathrm{g}}$) with the mirror cross-section function (of area $A_{\mathrm{m}}\cos\varphi$). The area of the resulting image would then be of order $A_{\odot,\mathrm{g}} + A_{\mathrm{m}}\cos\varphi$. Since $A_{\odot,\mathrm{g}} \gg A_m$, we neglect here the small correction due to the finite size of the mirror.

The effective radiometric quantities required to evaluate retinal hazards are calculated from the spectral corneal irradiance and radiance in Eqs.(15-16) by weighted integrals similar to the one for the illuminance and the luminance in (Eq. 6), respectively, but using the hazard-specific weighting functions provided by ICNIRP (2013), Fig. 1, instead of the photopic $V_\lambda$ function, Fig. 2.

When calculating accumulated long exposures from a mirror purposely oriented to reflect sunlight continuously towards an observer it may be necessary to take into account the time evolution of Eqs. (15-16) as the mirror moves in Earth orbit. The time-dependent terms are $\cos\varphi$, $d_{\mathrm{m}}$, and $M(z)$. Additionally, if the mirror illuminates the observer as a transient phenomenon, due to the latter being in the center of the ground path during a mirror slew operation, the exposure to the direct beam takes place during a time interval $\Delta t \approx \alpha_\odot \, \omega^{-1}$, where $\omega$ is the angular slew speed (rad $s^{-1}$), and where the eye pupil size and the changes in the Sun-mirror distance during the exposure time have been neglected.

## 6. Results

This section reports the results of some basic evaluations of the order of magnitude of the retinal hazards and glare of orbital mirrors. The parameters used in the calculations are listed below.

### 6.1 Main parameters

**Orbital mirror**: square mirror with side length $l_{\mathrm{m}} = 54$ m, area $A_{\mathrm{m}} = 2916$ $m^2$, orbit altitude $h_{\mathrm{m}} = 625$ km. Spectral reflectance $\rho_\lambda = 0.9$ (for simplicity assumed constant for all wavelengths). **For retinal hazard calculations:** Mirror located at the zenith of the observer, $z = 0°$, airmasses $M(0°) = 1$, distance to the observer $d_{\mathrm{m}} = h_{\mathrm{m}}$. Ground illuminated region of radius 2906.5 m, area 26.54 $km^2$. Angle between the normal to the mirror and the direction of the Sun $\varphi = 36.5°$. The largest dimension of the mirror seen from the observer subtends a linear angle $\alpha_{\mathrm{s}} = 0.30$ arcmin. The mirror cross section is 2344 $m^2$, and its solid angle $\Omega_{\mathrm{s}} = 6.00 \times 10^{-9}$ sr. Its magnitude in the Johnson V band (zero airmasses) seen by the observer is $-16.52$. Ground illuminance 8.7 lx, satellite luminance seen from ground $1.45 \times 10^9$ cd/$m^2$. **For glare calculations:** mirror anywhere in the upper hemisphere above the observer. Mirror angle $\varphi$ and distance to the observer $d_{\mathrm{m}}$ determined by the geocentric XYZ coordinates of the vertices of the triangle Sun-mirror-observer (section 6.4).

**Observer**: located at sea level, photopic pupil diameter $d_{\mathrm{p}} = 3$ mm. **For retinal hazard calculations**: looking directly at the space mirror ($\theta = 0°$, foveal fixation) with and without instruments; **For glare**: age $A_y = 31$ yr, eye pigmentation coefficient $p = 0$, unaided eye exposed to the direct irradiance of the mirror located at different angles $\theta$ within the field of view (section 6.4).

**Telescope** (for retinal hazard calculations)**:** aperture diameter $D_{\mathrm{T}} = 225$ mm, magnification $\beta = 75\times$, exit pupil diameter $d_{\mathrm{T}} = 3$ mm, spectral transmittance of the optics $\xi_{T,\lambda} = 0.9$ (for simplicity assumed constant for all wavelengths).

**Sun:** solar spectral irradiance at 1 au, $E_{\odot,\lambda}$, according to *sun_reference_STIS_002* from the CALSPEC database (STScI 2023, Bohlin et al. 2020). Magnitude in the Johnson V band (zero airmasses): −26.77. Sun illuminance: 133 700 lx (extra-atmospheric), 109 550 lx (on ground, Sun at zenith) Sun luminance: $1.97\times 10^9$ cd/m² (extra-atmospheric), $1.61\times 10^9$ cd/m² (seen from ground, Sun at zenith). Distance Sun-satellite $d_{\odot} = 1$ au. Sun angular diameter (at 1 au) $\alpha_{\odot} = 31.97$ arcmin, solid angle (at 1 au) $\Omega_{\odot} = 6.79\times 10^{-5}$ sr.

**Atmosphere**: Molecular (Rayleigh) optical depth at sea level $\tau_{\mathrm{R},\lambda} = 0.00879\ \lambda_{[\mu\mathrm{m}]}^{-4.09}$ (Teillet 1990). Aerosol optical depth at sea level and $\lambda$ =550 nm, $\tau_{\mathrm{A},\lambda=500} = 0.1$. Spectral dependence of $\tau_{\mathrm{A},\lambda}$ according to McComiskey et al. (2008), with Angström exponents for scattering and absorption equal to 1.

### 6.2 Retinal photochemical ("Blue light") hazard

Retinal photochemical hazards are evaluated in terms of the effective radiances (W m$^{-2}$ sr$^{-1}$) or irradiances (W m$^{-2}$) on the eye cornea, depending on the angular size of the source. These effective radiometric quantities are the integrals over wavelengths of the spectral radiances $L_{\mathrm{r},\lambda}$ (or irradiances, $E_{\mathrm{r},\lambda}$) weighted by the corresponding action spectra (Fig.1). Two action spectra are used, B($\lambda$) for general healthy adult population and A($\lambda$) for the aphakic one (people without sufficient short-wavelength filtering in the eye lens, as aphakic adults and very young infants). The results of these integrations are then compared with the maximum exposure limits, specific for each hazard type. The exposure limits generally depend on the angular size of the source and also on the exposure time. The photochemical limits are formulated for a pupil diameter of 3 mm, and shall be suitably modified if the observer has other effective pupil diameter. Detailed information for carrying out these calculations can be found in ICNIRP (2013).

Figure 3 shows the effective irradiance for blue light photoretinopathy, both for general (full blue line) and aphakic (dash-dotted blue line) populations, and the ICNIRP blue light irradiance exposure limit (red line), as a function of the exposure time. The three panels correspond to the unaided eye view of the Sun (left), the unaided eye view of the space mirror (center), and the space mirror viewed through a telescope, at 75× magnification with matched pupils (other parameters listed in subsection 6.1).

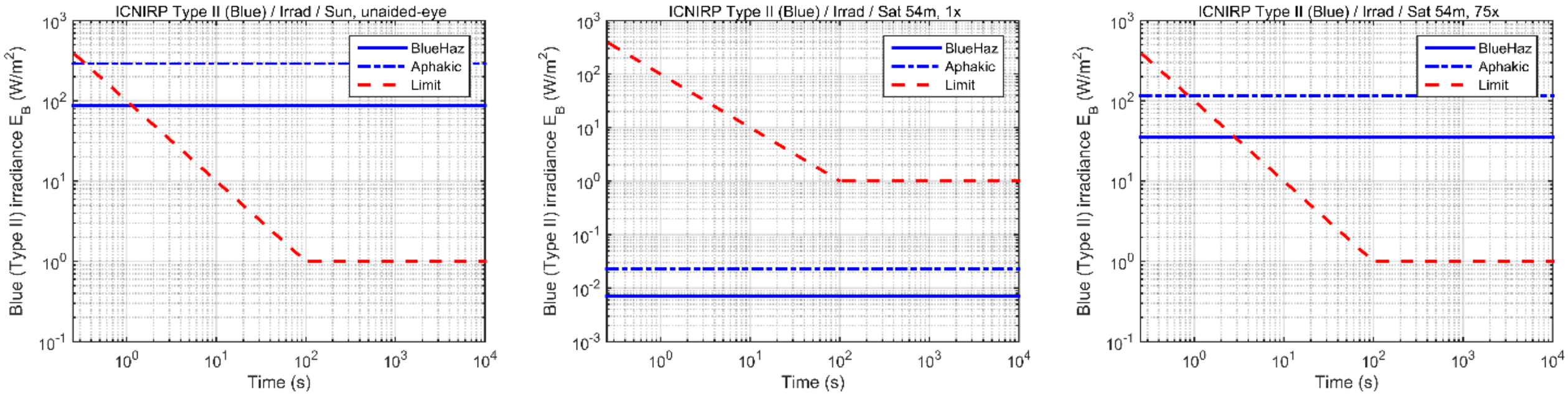

**Fig. 3.** Effective blue light irradiances (W m$^{-2}$) for the general population (full blue line) and aphakic population (dash-dotted blue line), and ICNIRP blue light irradiance exposure limit (dashed red line), versus exposure time in s. The light sources are: (left) the Sun, viewed with the unaided eye; (center) orbital mirror, unaided eye; (right) orbital mirror, seen through a telescope at 75× magnification with matched pupils. See parameters in subsection 6.1.

According to this figure, the blue light exposure limits for unaided eye Sun viewing are attained after ∼ 1 s exposure, for the general population, and ∼ 0.3 s for the aphakic one, the latter being close to the nominal reaction time of the aversion reflex (∼ 0.25 s). Unaided eye viewing the orbital mirror, under the conditions and parameters used here, does not seem to pose a remarkable risk, since the associate effective irradiance is about two orders of magnitude (or more) smaller than the red line limits. However, seeing the same object through the telescope, the exposure limits would be attained after ∼ 3 s (general population) and ∼ 0.8 s (aphakic). The difference between instrumental and unaided eye viewing is due to the larger size of the retinal geometric image of the source using a telescope.

### 6.3 Retinal thermal hazard

The hazard for thermally induced photoretinopathy is evaluated using the thermal weighting function R($\lambda$) shown in Fig. 1. The results obtained using the standard ICNIRP exposure limits are displayed in Figure 4, for unaided eye observation of the Sun (left panel), the orbital mirror (center), and for telescopic observation of the latter (right).

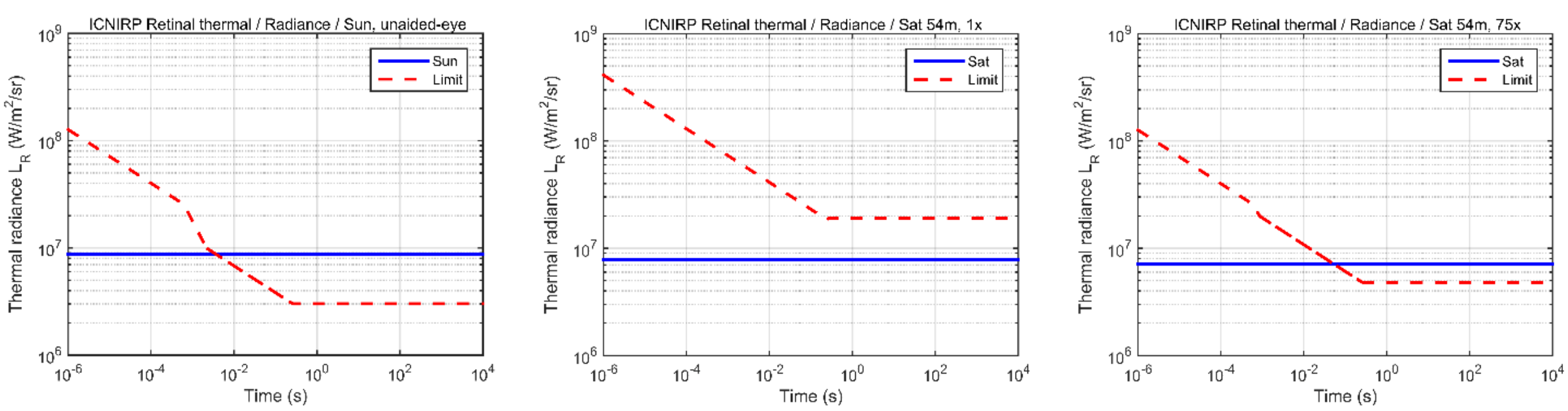

**Fig. 4.** Effective retinal thermal radiance (W m$^{-2}$ sr$^{-1}$) of the sources, seen from the observer location (full blue lines) and ICNIRP retinal thermal radiance exposure limits (dashed red lines), versus exposure time in s. The light sources are: (left) the Sun, viewed with the unaided eye; (center) orbital mirror,

unaided eye; (right) orbital mirror, seen through a telescope at 75× magnification with matched pupils. Additional parameters are listed in subsection 6.1.

Unaided eye view of the Sun (left, thermal radiance $8.77 \times 10^6$ W m$^{-2}$ sr$^{-1}$) would reach the exposure limit in ~ 3.7 ms. At a first sight these results could appear contradictory with the well-known fact that unaided eye view of the Sun, with normally constricted pupils in healthy adult eyes, does not produce thermal retinal burns, and the damage after several seconds of exposure is mainly photochemical (Vos and van Norren 2001). However, it should be kept in mind that ICNIRP exposure limits for thermal retinal hazard of sources emitting visible radiation (not excluding additional emissions in other spectral regions), shown in Fig. 4, are calculated under the assumption that the observer has a 7 mm diameter pupil during the first 0.25 s of exposure, followed by a constricted 3 mm diameter pupil afterwards. This assumption is relevant to protect observers in laboratory or workplace settings against incoherent light pulses, if their eyes are adapted to very low luminaces at the time of receiving the first flashes of light. While reasonable for these settings, it turns out to be a very conservative approach for usual Sun unaided eye viewing with photopically adapted pupils (ICNIRP 2013, Schulmeister 2013), like the situation considered in this paper (3 mm diameter from the beginning of the exposure). For a 3 mm pupil the thermal radiance limits should be increased by a factor $(7\ \mathrm{mm}/3\ \ \mathrm{mm})^2 = 5.4$, resulting in $16.4 \times 10^7$ W m$^{-2}$ sr$^{-1}$ for times longer than 0.25 s, which makes the Sun to remain in the safe side regarding retinal thermal exposure, although close to the limit (below by a factor ~ 2).

Unaided eye viewing of the mirror (Fig. 4, center) does not appear to pose an immediate retinal thermal hazard, even for long observation periods, although the ICNIRP exposure limit for times longer than 0.25 s ($1.90 \times 10^7$ W m$^{-2}$ sr$^{-1}$) is only 2.4 times larger than the thermal radiance produced at the eye cornea by the mirror ($7.89 \times 10^6$ W m$^{-2}$ sr$^{-1}$, which is smaller than the Sun one due to the mirror reflectance $\rho_\lambda = 0.9$). In this case the use of the original ICNIRP retinal thermal limits is well justified, since observing the sky at nighttime the eye pupil may be dilated at the instant of receiving the light from the mirror. Even in these conditions, the observation does not appear to present an hazard.

Telescopic views of the mirror (Fig. 4, right) would produce a slightly smaller corneal radiance than unaided eye viewing (due to the internal transmittance $\xi_{T,\lambda} = 0.9$ of the instrument), of about $7.10 \times 10^6$ W m$^{-2}$ sr$^{-1}$, attaining the radiance exposure limit after ~ 53 ms. Although the radiances are similar in both cases, the larger size of the telescopic retinal image of the mirror makes the difference. Here again it is in place the remark made above regarding exposure conditions. Even if the observer had a dilated pupil of 7 mm, the effective pupil for telescopic observations is the minimum of the observer's and the telescope exit pupil, which, in the conditions of our calculations, is 3 mm. The ICNIRP thermal limits should in this case be increased by a factor $(7\ \mathrm{mm}/3\ \ \mathrm{mm})^2$, which, for observation times of 0.25 s and longer, implies passing from $4.79 \times 10^6$ W m$^{-2}$ sr$^{-1}$ to $26.08 \times 10^6$ W m$^{-2}$ sr$^{-1}$, a factor 3.7 times above the actual exposure.

### 6.4 Glare

Disability glare may become a serious hazard when performing critical operations that require immediate responses, e.g., driving any kind of vehicles. For regulatory purposes, glare may be quantified by the percent threshold increment $f_{\mathrm{TI}}$ (CEN 2015), calculated using Eq. (10). According to this equation, the $f_{\mathrm{TI}}$ depends linearly on the veiling luminance, and non-linearly on the average luminance $L_{\mathrm{av}}$. The veiling luminance is additive, and the contribution of each source $k$ is given by $L_{\mathrm{veil}}(\theta_k) = \Psi(\theta_k)\, E_{\mathrm{v}k}$, i.e. the product of its corneal illuminance and the glare point spread function $\Psi(\theta)$ given by Eqs. (8)–(9), being $\theta$ the angular distance of the mirror with respect to the observer's line of sight.

The corneal illuminance $E_{\mathrm{v}k}$ produced by a space mirror, in turn, depends on various factors. One is the geometry of the Sun-mirror-observer system, which determines among other parameters the mirror attitude angle $\varphi$ for reflecting sunlight towards the observer (and hence the mirror cross section for capturing the Sun irradiance, and its solid angle seen from ground), as well as whether the mirror is illuminated by the Sun or is in the Earth's shadow. Other relevant factors are the reflectance of the mirror, the spectral transmittance of the terrestrial atmosphere, and the position angle of the mirror, $\theta$, within the observer visual field, via the $\cos\theta$ factor multiplying the normal illuminance (section 5).

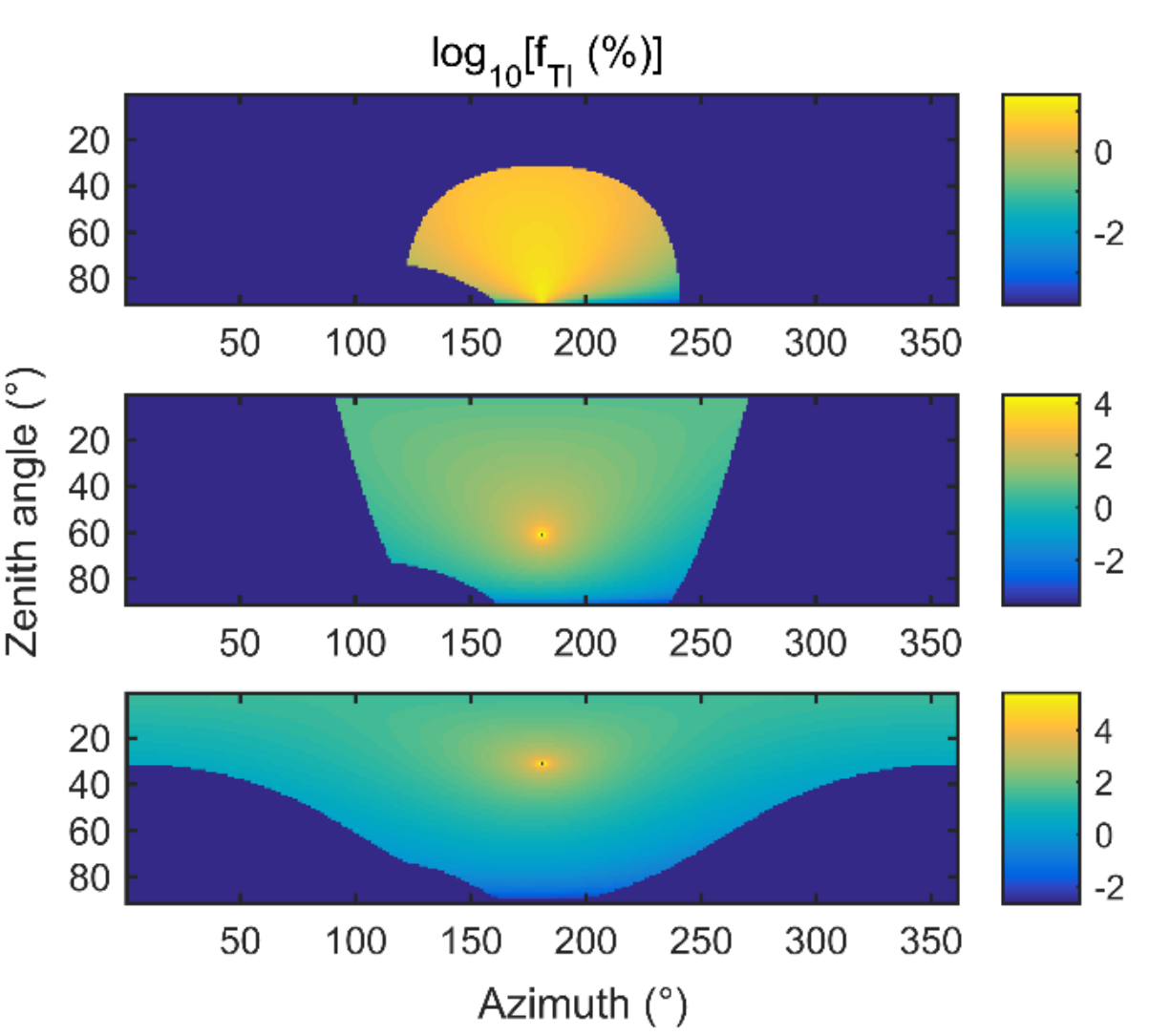

**Fig. 5.** Examples of the threshold increment $f_{\mathrm{TI}}$ (%) due to space mirrors located at the zenith angle - azimuth coordinates shown in the figures, for an observer whose line of sight points to azimuth 180° (due South) and altitude above the horizon 0° (upper panel), 30° (middle), and 60° (lower). Only satellites at angles $\theta$ smaller than 60° from the observer's line of sight are considered. The dark region close to the horizon around azimuth 120° corresponds to satellites in the Earth's shadow. Colorbar in logarithmic scale. See text for details.

To get some insights about the expected magnitude of the glare effects, Fig. 5 shows, in logarithmic color scale, the threshold increment $f_{\mathrm{TI}}$ (%) produced by a single space mirror with the characteristics listed in section 6.1, depending on its location in the upper hemisphere above the observer (zenith angles 0°–90°, and azimuths 0°–360°, in the observer's topocentric reference frame). The panels in Fig. 5 correspond to three directions of gaze, with lines of sight pointing at elevations 0°, 30°, and 60° above the horizon (from top to bottom, respectively) and azimuth 180° (South). The location of the observer is (45° N, 0° E), at sea level, and the subsolar point is located on the equator at (0° N, 108° E), being the Sun at a distance 1 au of the center of the Earth. Only the $f_{\mathrm{TI}}$ (%) values for space mirrors located at angular distances up to $\theta = 60°$ from the observer's line of sight are displayed, since this angular distance is the maximum one to include in the calculations according to the protocols of CEN (2015). The average luminance around the fixation point was set to $L_{\mathrm{av}} = 0.1$ cd m$^{-2}$. In order to assess satellite locations at very large zenith angles, the approximate airmass expression $M(z) \approx 1/\cos z$ was replaced in all the hemisphere by the more precise Eq. (13), which provides $M(90°) = 37.9$ airmasses at the horizon.

Depending on the position in the sky and the observer's line of sight, the values of $f_{\mathrm{TI}}$ (%) in Fig. 5 span several orders of magnitude, with wide sections of the visual field in which the presence of an isolated space mirror would increase the luminance contrast threshold by (much) more than the 10% to 15% permitted by the norms ($\log_{10}[f_{\mathrm{TI}}\ (\%)] = 1.00$ to 1.18). Since the veiling effects are additive, multiple mirrors above the observer's horizon shining simultaneously would increase significantly this hazard.

The plots of the veiling luminance $L_{\mathrm{veil}}$ produced by a single mirror have the same shape as those of Fig. 5, because $f_{\mathrm{TI}}$ (%) is directly proportional to it, as per Eq. (10). Since in this example the average luminance was taken as 0.1 cd m$^{-2}$, the veiling luminances would be $L_{\mathrm{veil}} = (0.1^{0.8}/65)\ f_{\mathrm{TI}}$ , in units cd m$^{-2}$. For a satellite located in a position for which $f_{\mathrm{TI}}\ (\%) = 10$ this would amount to $L_{\mathrm{veil}} = 0.024$ cd m$^{-2}$, about 120 times the luminance of a pristine sky of 22.0 Johnson V magnitudes per square arcsecond (200 µcd m$^{-2}$, Bará et al. 2020). These values can be significantly greater for mirrors closer to the observer's line of sight. Note that this intraocular scattered light adds to the atmospheric scattering described by Hainaut (2026) and Kocifaj et al. (2026), increasing the perceived skyglow, the only difference between them being the medium (atmosphere or eye structures) that sacatters the light before it arrives to the retinal photoreceptors.

## 7. Additional remarks and conclusions

The results of this paper show that sunlight reflected by a single orbital mirror, like the ones considered in some current space projects, poses a non-negligible risk of photochemical hazard for the retina if seen through optical instruments. This risk increases for vulnerable populations. Thermal effects in small telescopes (22.5 cm aperture, 75×) appear to be close to the exposure limits. Unaided eye viewing of a single mirror keeps the retinal hazards low. However, disability glare is expected to be unacceptably high, becoming a significant risk in critical visual operations requiring immediate awareness and response (e.g. driving).

The combined effects of the simultaneous presence of several space mirrors shining in the visual field of the observer, viewed with unaided eye or through optical instruments, will depend on the angular separation between them. Photochemical and thermal hazards will increase noticeably if several mirrors, seen from the observer, are enough close angularly as to be considered a single light source, pointlike or extended, according to ICNIRP criteria. Mirrors widely separated in the field of view, in turn, will affect the retina at different locations. Glare effects are additive, and the light from several mirrors shining simultaneously will in general be significantly more disabling than that from a single one.

Glare does not only affect visual performance during steady illumination, it is also expected to be a relevant issue during mirror slewing operations. Additional information on the number, spatial distribution, and angular slewing rate of the mirrors is needed in order to evaluate the effects of these transient flashes of light.

## Funding and Generative IA declaration

This work received no specific funding. Generative AI has not been used.

## Conflict of Interest

The authors declare that they have no competing interests.